\documentclass[a4paper,fleqn]{cas-dc}

\usepackage[numbers]{natbib}
\usepackage{tabu} 
\usepackage{xcolor}
\usepackage{gensymb}
\usepackage{float}
\usepackage{blindtext}
\usepackage{multicol}
\usepackage{fixltx2e}
\usepackage[center]{caption}

\begin{document}
\let\WriteBookmarks\relax
\def\floatpagepagefraction{1}
\def\textpagefraction{.001}
\shorttitle{Shadow Bands}
\shortauthors{Madhani et~al.}

\title [mode = title]{Observation of Eclipse Shadow Bands Using High Altitude Balloon and Ground-Based Photodiode Arrays}

\author[1]{Janvi P. Madhani}[type=editor,
                        auid=000,bioid=1,
                        orcid=0000-0002-0913-991X]
                        
\cormark[1]
\fnmark[1]
\ead{jmadhan1@jh.edu}

\address[1]{Department of Physics and Astronomy, University of Pittsburgh,
Pittsburgh, PA. 15260, USA.}

\author[1]{Grace E. Chu}

\author[2]{Carlos Vazquez Gomez}

\address[2]{School of Engineering, University of Pittsburgh,
Pittsburgh, PA. 15260, USA.}

\author[2]{Sinjon Bartel}

\author[1]{Russell J. Clark}

\author[1,3]{Lou W. Coban}

\address[3]{Allegheny Observatory, 159 Riverview Ave, Pittsburgh, PA 15214, USA.}

\author[2]{Marshall Hartman}

\author[1,3]{Edward M. Potosky}

\author[1]{Sandhya M. Rao}

\author[1,3]{David A. Turnshek}

\cortext[cor1]{Corresponding author}

\begin{abstract}
The results of an investigation into whether or not eclipse shadow bands have an atmospheric origin are presented. Using high altitude balloon and ground-based photodiode arrays during the 21 August 2017 total solar eclipse, data revealing the light patterns before and after totality were collected at 600 Hz. These data were then analyzed using spectrograms, which provide information on intensity fluctuations in the frequency space time domain. Both at the altitude of the balloon ($\sim 25$ km) and on the ground, a sustained $\sim 4.5$ Hz  signal was detected a few minutes before and after totality. This signal was coherent over a scale greater than 10 cm and detected in four separate balloon photodiodes and 16 separate ground photodiodes. At higher frequencies, up to at least 30 Hz, brief chaotic signals that were disorganized as a function of time were detected on the ground, but not at the altitude of the balloon. These higher frequency signals, which we attribute to atmospheric scintillation, appeared mostly uncorrelated over a length scale of 10 cm. Some of our ground arrays utilized red and blue filters, but neither the sustained 4.5 Hz signal nor the chaotic higher frequency signals showed a strong dependence on filter color. On the ground we made a video of the shadow bands on a scaled white screen. We judged that the bands were roughly parallel to the orientation of the bright thin crescent Sun before and after totality. Thus, if the $\nu \approx 4.5$ Hz frequency signal is identified with the peak-to-peak shadow band wavelength of $\lambda \approx 13$ cm measured in the video, it can be inferred that their propagation velocity was about $v \approx 59$ cm s$^{-1}$ ($\approx 2.1$ km hr$^{-1}$). Shadow band signals other than the sustained signal at $\sim 4.5$ Hz are consistent with atmospheric scintillation theory.

These results are surprising. Based on accounts in the literature we expected to confirm the atmospheric scintillation theory of eclipse shadow bands, but instead we detected a sustained $\sim 4.5$ Hz signal at both high altitude and on the ground, consistent with the type of shadow band signal visual observers often report before and after totality. This signal cannot be due to atmospheric scintillation and we ran a check to make sure this signal is not an artifact of our electronics. We recommend that additional searches for eclipse shadow bands be made at high altitude in the future.  
\end{abstract}

\begin{keywords}
eclipse shadow bands \sep total solar eclipse \sep atmospheric scintillation 
\end{keywords}

\maketitle

\section{Introduction}

During the few minutes before and after totality of a solar eclipse, indistinct bands of light and dark, known as shadow bands, appear to sometimes chaotically undulate across the surface of the Earth. Observers have reported various results. For example, Quann \& Daly (1972) \citep{1972JATP...34..577Q} report that the bands  appear to be parallel to one another but have different widths and move at different speeds.
Marschall et al. (1984) \citep{Marschall:84} estimate that the intensity difference between the light and dark portion of each band is as much as 2-3$\%$, which would make them elusive to measure without a proper setup. As totality approaches, the theoretical study of Codona (1986) \citep{1986A&A...164..415C}, which assumes shadow bands have an atmospheric origin, indicates that the spacing between bands should decrease and the contrast between bands should increase closer to totality, making them easier to distinguish with the naked eye. These are among the more detailed results on shadow bands that have been reported over the last 50 years.

The earliest observers appear to have accounted for them as some sort of interference pattern similar to the Fraunhofer patterns produced by a sharp edge when illuminated by a point source (e.g., see the background historical discussions in Quann \& Daly 1972 \citep{1972JATP...34..577Q} and Marschall et al. 1984 \citep{Marschall:84}). Also, in a series of reports published between 1938 and 1948, Feldman suggested that some type of diffraction-interference effect was both viable and likely (e.g., see Feldman 1940 \citep{1940PA.....48..182F}). Pickering \& Pickering (1890) \citep{1890AnHar..18...85P} were the first to suggest that shadow bands have an atmospheric origin. 

Indeed, the most prominent modern-day theory suggests that turbulent density fluctuations in Earth's atmosphere act as a lens to rapidly differentially refract light from the thin crescent of the eclipsed Sun as it reaches the surface of the Earth (e.g., Codona 1986 \cite{1986A&A...164..415C}).
This is similar to scintillations arising in Earth's turbulent atmosphere (Osborn et al. 2015 \citep{2015MNRAS.452.1707O}), in the same manner that turbulence in the atmosphere is the cause of the rapid intensity fluctuations which make stars twinkle when viewed with the unaided eye.
Shadow bands are now commonly theorized to be a phenomenon resulting from this atmospheric scintillation. Since the cells of atmospheric turbulence and wind velocity components are different for each eclipse,  the movement and intensity of shadow bands cannot be precisely predicted without atmospheric data.
Thus, according to atmospheric scintillation theory, precise shadow band patterns are different for each eclipse. They also depend on the duration of the eclipse and the location of the observer from the center-line of the path of totality. Yet, in order to show that this theory is viable and consistent with observations, one would have to demonstrate that shadow bands {\it are not} detectable above the turbulent layer of Earth's atmosphere. 

On 21 August 2017, during the total solar eclipse that took place over the continental United States, we flew a high altitude balloon above Barren Plains, Tennessee (near the eclipse centerline), to image the solar eclipse and the Moon's shadow on Earth, and also to search for shadow bands from near space, i.e., well above the Earth's turbulent atmosphere.\footnote{See \href{http://www.youtube.com/watch?v=mvh3tRE6uNk}{www.youtube.com/watch?v=mvh3tRE6uNk} for a video which summarizes our balloon flight.} Observations near totality took place at an altitude of $\sim 25$ km, which is above the tropopause in the lower stratosphere.  Our high altitude balloon was equipped with a light sensor payload consisting of an array of four photodiodes. In addition, similar data were collected on the ground so that results above and below the atmosphere could be compared. If shadow bands were detected above the turbulent atmosphere, this would clearly invalidate any theory that postulates eclipse shadow bands solely have an atmospheric origin. To our knowledge there are no prior results on searches for eclipse shadow bands above Earth's turbulent atmosphere. While our ground-based photodiode arrays did detect patterns that were consistent with chaotic shadow bands due to atmospheric scintillation, no such patterns were detected by our balloon arrays. However, we also detected a persistent signal at $\sim 4.5$ Hz in both our balloon and ground photodiode arrays, indicating that some component of the observed shadow bands originates above the turbulent atmosphere.    

In the next section (\S\ref{theory}), 
the atmospheric scintillation theory for shadow bands is described in greater qualitative detail. Other relevant issues involving diffraction and/or interference effects are noted in \S\ref{other}. The instrumentation and measurement process developed to collect data is described in \S\ref{instrumentation}. This includes the set-ups for the near space measurements from the high altitude balloon as well as for the ground-based measurements. 
The data processing and analysis is presented in \S\ref{data}. Our main analysis tool was the spectrogram, which provides information on intensity fluctuations in the frequency space time domain. A discussion and some conclusions are presented in \S\ref{summary}.

\section{Qualitative Theory of Atmospheric Scintillation} \label{theory}

The leading theory for explaining the origin of eclipse shadow bands is that they are generally an effect of atmospheric scintillation (e.g., Quann \& Daly 1972 \citep{1972JATP...34..577Q}, Marschall et al. 1984 \citep{Marschall:84}, and Codona 1986 \citep{1986A&A...164..415C}). 
An observer looking at a distant point source star will see the starlight refracted as it travels through Earth's turbulent atmosphere. The first-order location of the star's refracted image in the sky depends on the star's zenith angle and the refractive index of the atmosphere, which varies with altitude. The second-order correction to the star's location depends on the details of atmospheric turbulence. The turbulence is caused by rapid movement of air regulated by atmospheric temperature and pressure. This causes an observer to see a star as twinkling rather than see it as a steady point of light. This effect is known as scintillation; the more popular term is ``astronomical seeing.'' For objects that subtend a much larger angular size (e.g., a planet), incoming light is usually not differentially refracted by turbulence more than the object's angular size, even if the object is closer to the horizon and passing through more atmosphere. Something similar holds for sunlight reaching the Earth during a total solar eclipse. The light from the Sun is reduced to a thin crescent as the eclipse nears totality. Although the crescent-shaped slit of light is not an unresolved point source, it acts like a collection of point sources, and the incoming light from this elongated source is what is refracted.

A good description of what is happening can be found in shadowgraphy, the technique used to make a shadowgram (Settles 2001 \citep{Settles}). This is an optical method that reveals non-uniform regions in a transparent medium. It is simpler than a schlieren photography system because it does not require a knife edge. It is often used to visualize flows. Since disturbances in a transparent medium refract light waves, the disturbances can effectively cast shadows, and a shadowgram can be used to see differences in, for example, temperature, density, pressure, or a shock wave moving through transparent air. Similarly, if one were to set up a light source and a screen with a heat source in between and below them,  the motion of the air above the heat source would be seen via the shadow it casts on the white screen. If the bright source is thin and elongated, the shadows which are cast will generally be aligned parallel to the orientation of the source. It is important that the distance between the disturbance in the transparent medium and the shadow cast on the screen is not too large relative to the distance of the light source, but of course this holds for the case of a total solar eclipse.
In a shadowgram the differences in light intensity are proportional to the second spatial derivative (Laplacian) of the refractive index field in the transparent medium. 

Thus, when the background light source is the bright thin crescent of the Sun before and after totality, shadows cast by Earth's turbulent atmosphere being carried by upper atmospheric winds, will scintillate and appear as light and dark bands on the ground which are parallel to crescent.    
The theoretical study of Codona (1986) \citep{1986A&A...164..415C} provides detailed predictions about the characteristics of shadow bands in the context of the atmospheric scintillation theory, but the theory has been impossible to test in detail due to the lack of high-quality atmospheric data during an eclipse.

\begin{figure*}[pos=h]
    \centering
    \includegraphics[width=.9\linewidth]{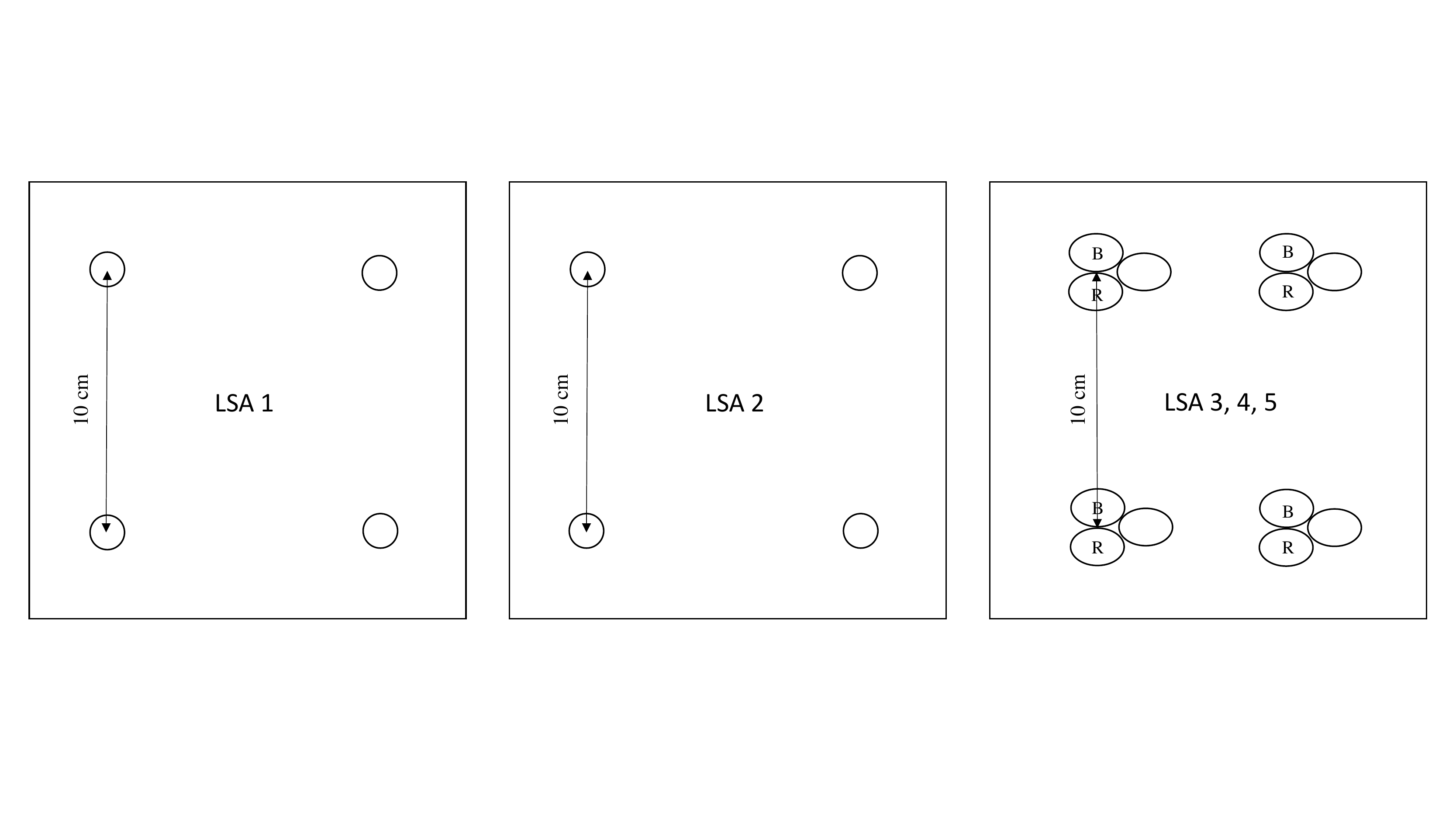}
    \caption{The five Light Sensor Arrays (LSAs), each with four photodiode channels per array. LSA 1 (left) was on the high altitude ballon}
    \label{fig:fig1}
\end{figure*}

\begin{table*}[width=.9\linewidth,pos=h]
\centering
\begin{tabular}{ |c||c|c|c|  }
 \hline
 \multicolumn{4}{|c|}{Light Sensor Setup with 4 Photodiode Channels per Array} \\
 \hline
 Light Sensor Array (LSA) Number & Location & Wavelength Range (color) & Pointed at\\
 \hline
 \hline
 1 & Balloon & 320 - 1000 nm (clear) & Approximately Zenith \\
 \hline
 2 & Ground & 320 - 1000 nm (clear) & Zenith \\
 \hline
 3 & Ground &  320 - 730 nm (clear) & Sun \\
 \hline
 4 & Ground & ~ 400 - 500 nm (blue) & Sun \\
 \hline
 5 & Ground & ~ 550 - 800 nm (red) & Sun \\
 \hline
\end{tabular}
\end{table*}
\label{table:table1}

\section{Other Effects Relevant to Shadow Bands} \label{other}

Past explanations for shadow bands have involved discussions of diffraction and interference theory.  
The rotating-Earth issue has often been given as a reason for questioning the idea that any type of diffraction-interference pattern could be detected a few minutes before and after a total solar eclipse. For example, given our latitude of $36.6 \degree$N in Tennessee and the Sun's $\sim 26 \degree$ zenith angle near totality, the surface of the Earth would be moving under the umbra of the Moon's shadow at a speed of $\sim 3.4 \times 10^4$ cm s$^{-1}$. In principle this would affect the observed band pattern.
However, any diffraction-interference pattern from an eclipse would be dynamic due to the fact that the Moon and Sun are both moving at different angular rates in the sky relative to the surface of the Earth.  
Making a detailed prediction about what kind of diffraction-interference pattern might be observed is beyond the scope of this paper. 
However, we can summarize a few ideas about some effects that might be relevant. 

To obtain strong interference patterns a monochromatic light source is required.  
Chromospheric emission from the Sun becomes more easily visible in the moments before and after totality when a large fraction of the photosphere is occulted. The chromosphere has strong emission lines especially due to hydrogen and helium.\footnote{To visualize this see, for example, the chromospheric flash spectrum taken by Yujing Qin during the August 2017 total eclipse at \href{http://apod.nasa.gov/apod/ap170907.html}{apod.nasa.gov/apod/ap170907.html}.} Therefore, any calculation of interference would need to model multiple sources of monochromatic emission at the wavelengths of chromospheric emission lines around a circle or arc-like region surrounding the  Moon as a function of time.

The so-called Lloyd's mirror effect might be an important mechanism for producing shadow bands. This has been outlined by Stanford (1973) \citep{1973AmJPh..41..731S}. He emphasizes that any object, such as a cloud in the Earth's atmosphere which lies outside the geometric shadow of the Moon, could serve as a Lloyd's mirror from which light could be reflected back toward the Earth. The direct beam and reflected beam could then combine to produce interference. 
There could be more than one reflecting surface. And, for example, the dark surface of the Moon itself might act as the reflecting surface needed to produce shadow bands. The existence of this reflection is commonly observed during a crescent Moon phase when light is reflected from Earth to the Moon and back to Earth in a phenomenon known as Earthshine. 

Another effect that should be considered is the umbral pattern produced by an occulter (i.e., the Moon) as it eclipses the extended image of the Sun. When the Moon acts as the occulter it becomes a natural coronagraph, but nothing minimizes diffraction as is the case when using the Lyot and more advanced artificial coronagraph designs. The diffraction which will be seen in the umbral pattern involves a convolution of the Fresnel diffraction pattern and the extended image of the Sun, which is not easily computed (e.g., see Aime 2013 \citep{2013A&A...558A.138A}).

\begin{figure*}[pos=h]
    \centering
    \includegraphics[width=.9\linewidth]{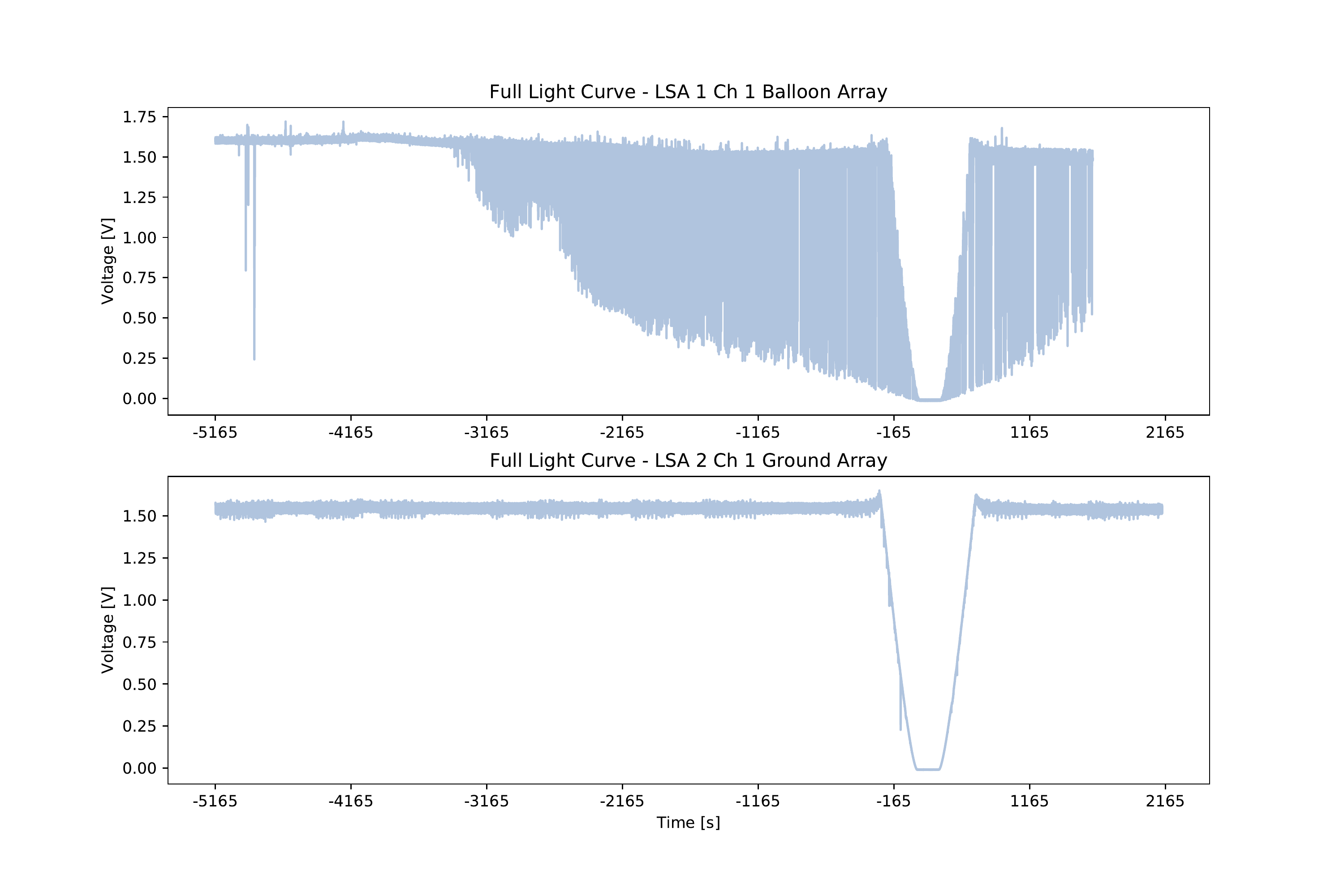}
    \caption{Examples of full light curves from two of the 20 photodiode channels. The top light curve is from photodiode 1 of LSA 1. The balloon was launched at a time near $-3200$ sec. Three ropes connecting sub-payloads in the payload string could occult the photodiodes throughout the flight. Therefore, the jagged nature of the light curve after launch is due to the spinning of the balloon payload, and was expected.  The spinning was typically at $\sim 0.1$ Hz, which means the frequency of the rope occultations was $\sim 0.3$ Hz. This spinning did not affect our search for $> 1$ Hz shadowbands above the atmosphere. The bottom light curve is from photodiode 1 of LSA 2, which was the ground control. In both light curves the signal appears low relative to its maximum for only $\sim 800$ sec, even though the partial phases of the eclipse lasted much longer.  This is because the photodiode signals were effectively saturated during most of the partial eclipse phases.}
    \label{fig:fig2}

\end{figure*}

\section{Instrumentation} \label{instrumentation}
For ground based measurements near totality, a white screen with a zenith angle of $\sim 26 \degree$ was set up so that its surface was perpendicular to the direction of the Sun at totality. This screen was primarily used for naked eye observations of shadow bands, and their presence was confirmed by many observers as well as on a video we obtained.

We utilized five Light Sensor Arrays (LSAs), each consisting of four photodiode channels. They were arranged in a square configuration with separations of length 10 cm (see Figure \ref{fig:fig1}).

The first two arrays (LSA 1 and LSA 2) used the Hamamatsu S1226-18BK silicon photodiode. Its  spectral response range of $\sim 320-1000$ nm peaked at $\sim 720$ nm. Though this spectral range was larger than necessary, we chose this photodiode for its ability to operate at very low temperature in near space as well as its ability to detect low light levels.  
LSA 1 was flown on the high altitude balloon; it was generally pointed toward zenith, but the swaying motion of the balloon payload meant that its exact pointing must have slowly varied near the time of totality. LSA 2 was placed on the ground near the white screen. The configuration of LSA 2 was identical to LSA 1, also pointing toward zenith. 

The three other ground-based arrays (LSA 3, LSA 4, and LSA 5) were all mounted on the same board, but were inclined to point toward the Sun at the time of totality, just like the white screen. They all used Hamamatsu S1087 photodiodes.
This particular photodiode was chosen for its low-light-level sensitivity and its spectral response range of $\sim 320-730$ nm, but its operating temperature was not a concern.
The LSA 3 photodiodes were unfiltered (clear), so they simply had the native spectral response of the Hamamatsu S1078.
LSA 4 used a blue filter with a spectral response that peaked at $\sim 450-490$ nm, while LSA 5 used a red filter with a spectral response that peaked at $\sim 635-700$ nm. The colored filters were used  to qualitatively assess if shadow bands had any characteristics that were color dependent.

Each of the five LSAs were connected to their own circuit, whose analog output was amplified, digitized, and then fed into a Raspberry Pi computer where it was stored as a voltage reading. The data were collected at 600 Hz, starting approximately an hour before totality and continuing after fourth contact for an additional twenty minutes. LSA 1 was powered by battery, whereas LSA 2, LSA 3, LSA 4, and LSA 5 used ground power, which introduced a $\sim 60$ Hz signal that could mostly be filtered out, as needed. 
Figure \ref{fig:fig1} and \ref{table:table1} summarize the configurations of the five LSAs.

\section{Data Analysis} \label{data}

\begin{figure*}[pos=h]
    \centering
    \includegraphics[width=.9\linewidth]{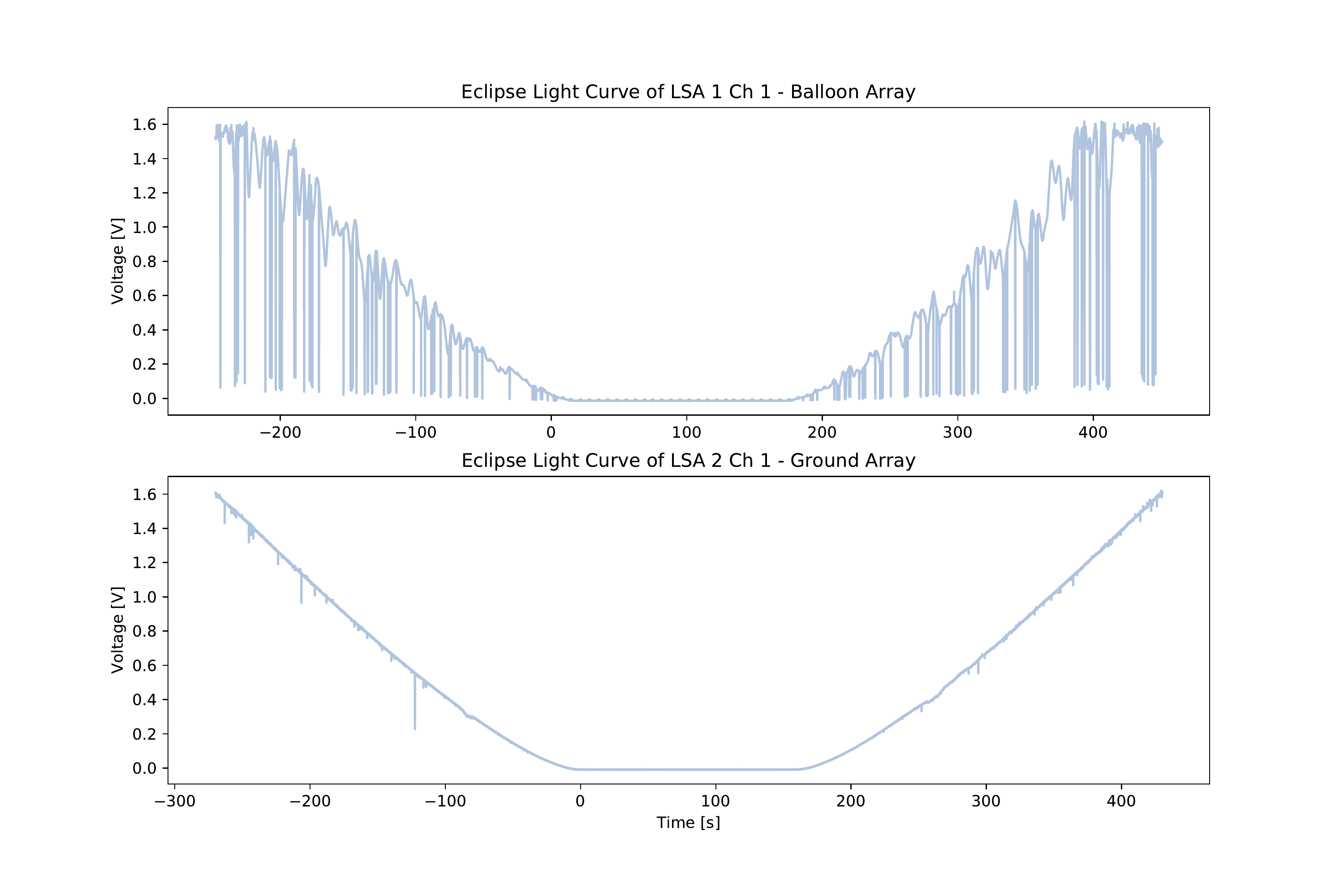}
    \caption{Expanded LSA 1 (top) and LSA 2 (bottom) light curves corresponding to Figure \ref{fig:fig2}, but truncated in time to be centered on the $\sim 700$ sec near mid-totality. Note the horizontal offset of the two graphs, since totality began at slightly different times for LSA 1 and LSA 2.}
    \label{fig:fig3}
    
\end{figure*}

Examples of raw data are presented in \S5.1, followed by our spectrogram analysis in \S5.2. For all of the photodiode channels, time is synchronized and defined so that time zero is the start of totality on all LSA light curves.  Totality lasted approximately 165 seconds. Note that there is a small offset between the beginning of totality for the balloon array (LSA 1) and the four ground arrays (LSA 2, 3, 4, 5) because the balloon was not exactly above our location during totality. During totality it was $\approx 11.2$ km, $72 \degree$ south of east, from the launch site where the ground LSAs were located, but still close to the center-line of totality. 

\subsection{Raw Data}
The top and bottom panels of Figure \ref{fig:fig2} depict the raw light curves obtained from photodiode 1 of balloon LSA 1 and photodiode 1 of ground LSA 2, respectively. Figure \ref{fig:fig3} shows the same data shown in Figure \ref{fig:fig2}, but over a shorter time interval centered on mid-totality. These are good examples of the light curves from all 20 photodiodes. Their similarities indicate that the data we collected are robust, and it is therefore sufficient to show only two examples.  

\subsection{Spectrogram Analysis}

After experimenting with a variety of methods to quantify any shadow band signals, we decided to use spectrograms. Spectrograms provide a way to visually represent the frequencies present in a sample of data as they vary over time. They can be computed by rolling a 2D Fast Fourier Transform (FFT) window over the length of a sample. The result then allows us to visualize evidence for frequency variations as a function of time. As with the light curves, spectrograms are synchronized so that totality in a particular LSA starts at time zero. 

\subsubsection{Steps Used to Form Spectrograms}

To form spectrograms for the stationary ground arrays (LSA 2, 3, 4, 5) we: (1) normalized each single photodiode light curve using a polynomial fit, (2) applied a 3$\sigma$-clipping algorithm to each light curve to remove spurious noise, (3) un-normalized each light curve using the same polynomial fit, (4) computed a spectrogram for each single-photodiode light curve, and (5) averaged four spectrograms to obtain an average spectrogram for each LSA. A single-photodiode spectrogram formed in step 4 would best represent the results for signals that were coherent over length-scales $<10$ cm, while the average formed in step 5 would best represent the results for signals coherent over length-scales $>10$ cm.  

\begin{figure*}[pos=h]
    \centering
    \includegraphics[width=.9\linewidth]{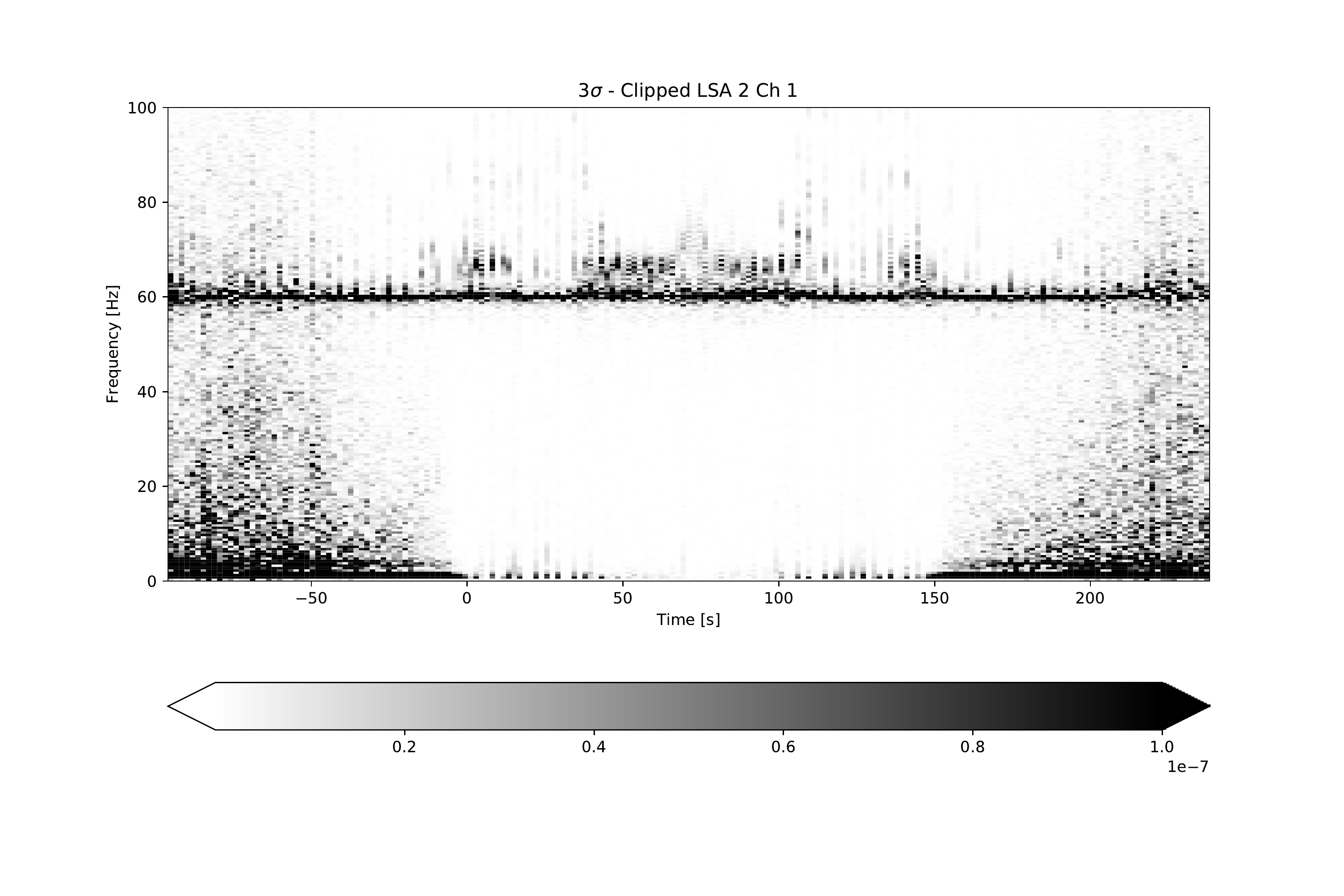}
    \caption{Example of a spectrogram formed using one of the 16 ground photodiodes. This shows the spectrogram formed using photodiode 1 of LSA 2 covering $\sim 100$ sec before totality to $\sim 100$ sec after totality. See \S5.2.1 for the procedure used to form a single-photodiode spectrogram. This shows frequencies up to 100 Hz. Note the signal near $\sim 60$ Hz due to our use of AC power for our ground arrays. However, there is no very useful signal to interpret above $\sim 30$ Hz. The light signals fade away but the $60$ Hz signal does not. The signals somewhat above 60 Hz during totality are also electronic noise.}
    \label{fig:fig4}
\end{figure*}

However, to form spectrograms for the high altitude balloon array (LSA 1), an additional step is needed. This is because the ropes connecting the sub-payload packages can occult single photodiodes as the payload spins. Overall the spinning introduces an $\sim 0.3$ Hz signal, which is below any signal of interest (see the Figure 2 caption). These occultations are usually extremely brief. For LSA 1 we account for the rope occultations before step 5 (above) by applying a bad-data-quality array to mask out bad regions in each of the four single-photodiode spectrograms before computing the average spectrogram.  Thus, the single-photodiode spectrograms from LSA 1 are unusable in certain time intervals, but the average spectrogram used to search for coherent signals over length scales $>10$ cm is improved due to the use of the data quality masks. 

\subsubsection{Results from Ground LSAs}

\begin{figure*}[pos=h]
    \centering
    \includegraphics[width=.9\linewidth]{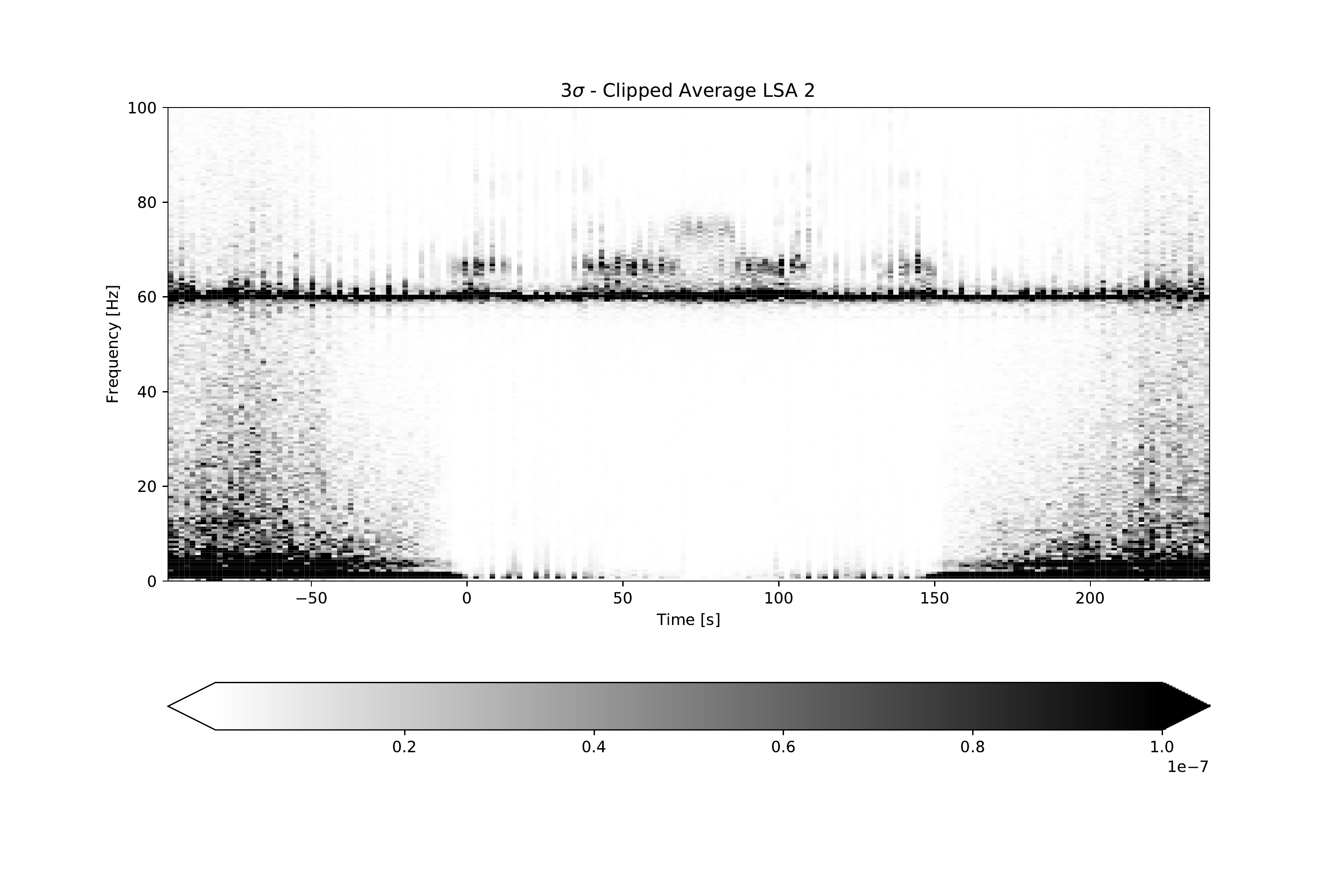}
    \caption{Example of an average spectrogram formed using four photodiodes from one of the LSAs. This is the average spectrogram for LSA 2 covering from $\sim 100$ sec before totality to $\sim 100$ sec after totality. Again, see \S5.2.1 for the procedure used to form an average spectrogram. Again, this shows frequencies up to 100 Hz. And again, note the signal near $\sim 60$ Hz due to our use of AC power for our ground arrays. However, there is no very useful signal to interpret above $\sim 30$ Hz. The light signals fade away but the $60$ Hz signal does not. The signals between 66 Hz and 76 Hz during totality are also electronic noise.}
    \label{fig:fig5}
\end{figure*}

The spectrogram from photodiode 1 of LSA 2 is shown in Figure \ref{fig:fig4} as an example of a single-photodiode spectrogram from a ground LSA. It covers from about 100 sec before totality to about 100 sec after totality and extends up to 100 Hz. Figure \ref{fig:fig5} shows the average spectrogram for LSA 2. The signal near $\sim 60$ Hz in both figures is due to our use of AC power for our ground arrays. Fortunately the spectrogram allows us to visually isolate this signal and search for signals at other frequencies. This $\sim 60$ Hz signal is visible in all of our ground-based spectrograms. Note that this $\sim 60$ Hz signal does not fade away during totality, but all lower frequency signals do fade away. This characteristic is observed in all of our ground-based LSAs. 

\begin{figure*}[pos=h]
    \centering
    \includegraphics[width=.9\linewidth]{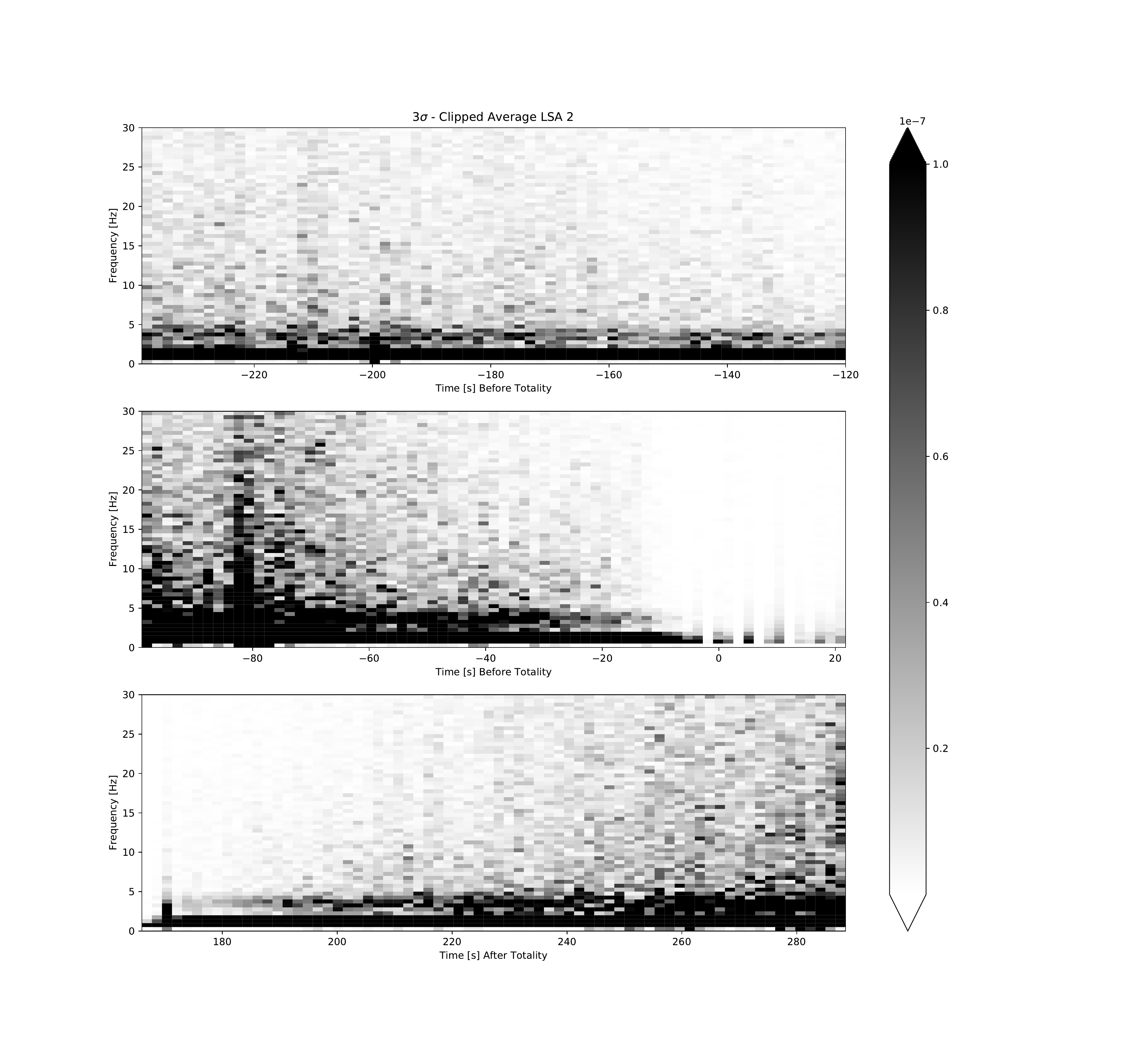}
    \caption{LSA 2 ground-based average spectrogram formed from the combined light curves of photodiodes 1, 2, 3, and 4. The spectrograms are shown in three time sections: about 4 to 2 minutes before totality, about 2 before totality, and about 2 minutes after totality. The chaotic signals at frequencies $> 5$ Hz and the sustained signal at $\sim 4.5$ Hz are apparent during all these times, but they all fade away during totality.}
    \label{fig:fig6}
\end{figure*}

Another general feature of all 16 single-photodiode spectrograms and the 4 average spectrograms (from LSAs 2, 3, 4, 5) is that above 5 Hz the signal is 
chaotic in nature, with few frequencies persisting for a duration longer than a few seconds. These chaotic signals in any one of the ground-based LSAs do not appear to be well correlated when comparing individual signals from the four photodiodes, but they are so ubiquitous that taking an average of them does not make them disappear (e.g., compare Figures \ref{fig:fig4} and \ref{fig:fig5}). Also, there are generally no very useful signals to interpret above $\sim 30$ Hz, although they probably exist, so we will restrict the next three spectrogram figures to have a maximum frequency scale of $30$ Hz. 

\begin{figure*}[pos=h]
    \centering
    \includegraphics[width=.9\linewidth]{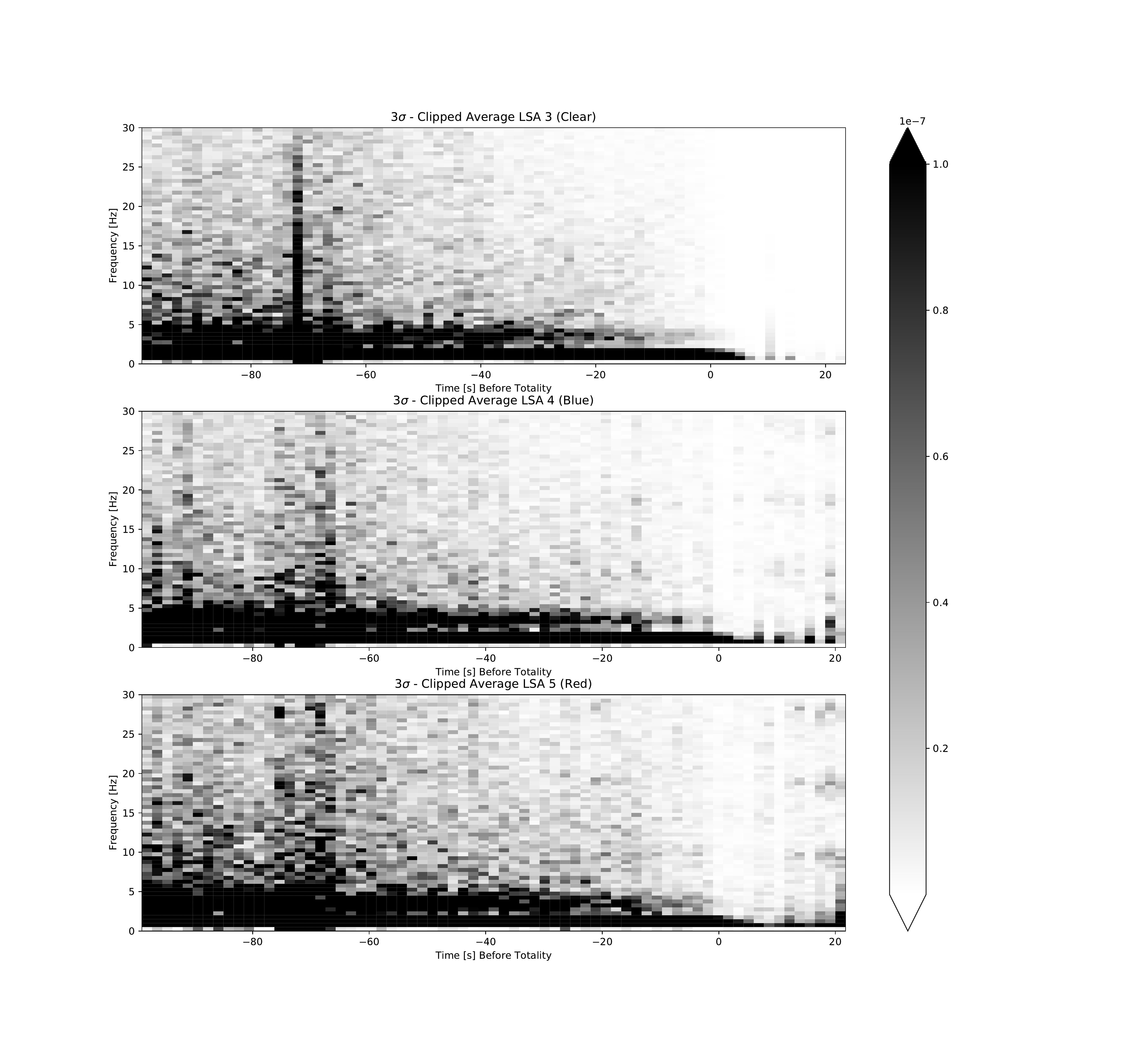}
    \caption{LSA 3 (clear), LSA 4 (blue filtered), and LSA 5 (red filtered) ground-based average spectrogram, from top to bottom, respectively, formed from the combined light curves of photodiode 1, 2, 3, and 4 in the LSA. The spectrograms are shown starting from about 2 minutes before totality. No obvious color dependence is present. The chaotic signals at frequencies $>5$ Hz and the sustained signal at $\sim 4.5$ Hz are apparent during all these times, but they all fade away as totality approaches.}
    \label{fig:fig7}
\end{figure*}

Figures \ref{fig:fig6} and \ref{fig:fig7} show the average spectrograms for zenith-pointing LSA 2 and sun-pointing LSAs 3 (clear), 4 (blue filtered), and 5 (red filtered). Figure \ref{fig:fig6} shows three sections for LSA 2: about 4 to 2 minutes before totality, about 2 minutes before totality, and about 2 minutes after totality. Figure \ref{fig:fig7} shows about 2 minutes before totality for LSA 3, LSA 4, and LSA 5. These all show the chaotic signals at frequencies $> 5$ Hz, especially when the frequencies are $<20$ Hz, but they fade away as totality approaches. These types of signals are what one might expect based on the atmospheric scintillation theory of shadow bands (\S2). 

Below 5 Hz, the spectrograms show a different character. A sustained $\sim 4.5$ Hz signal is present in all spectrograms. After the eclipse we checked to make sure this was not an artifact of our electronics at low light levels. We could not reproduce it.  It appears in all four photodiodes signals for all LSAs (20 signals in total). Moreover, it fades away as totality approaches and reappears after totality. In this respect, the $\sim 60$ Hz contamination was helpful in being able to differentiate between an electronic signal not due to light detected by the photodiodes and an apparently real optical signal caused by light. While the electronic signal persists even during totality, the $\sim 4.5$ Hz signal fades out as totality nears and then returns after totality.  This $\sim 4.5$ Hz signal is clearly coherent on scales larger than the 10 cm spacing of the photodiode channels in a LSA. 

\begin{figure*}[pos=h]
    \centering
    \includegraphics[width=.9\linewidth]{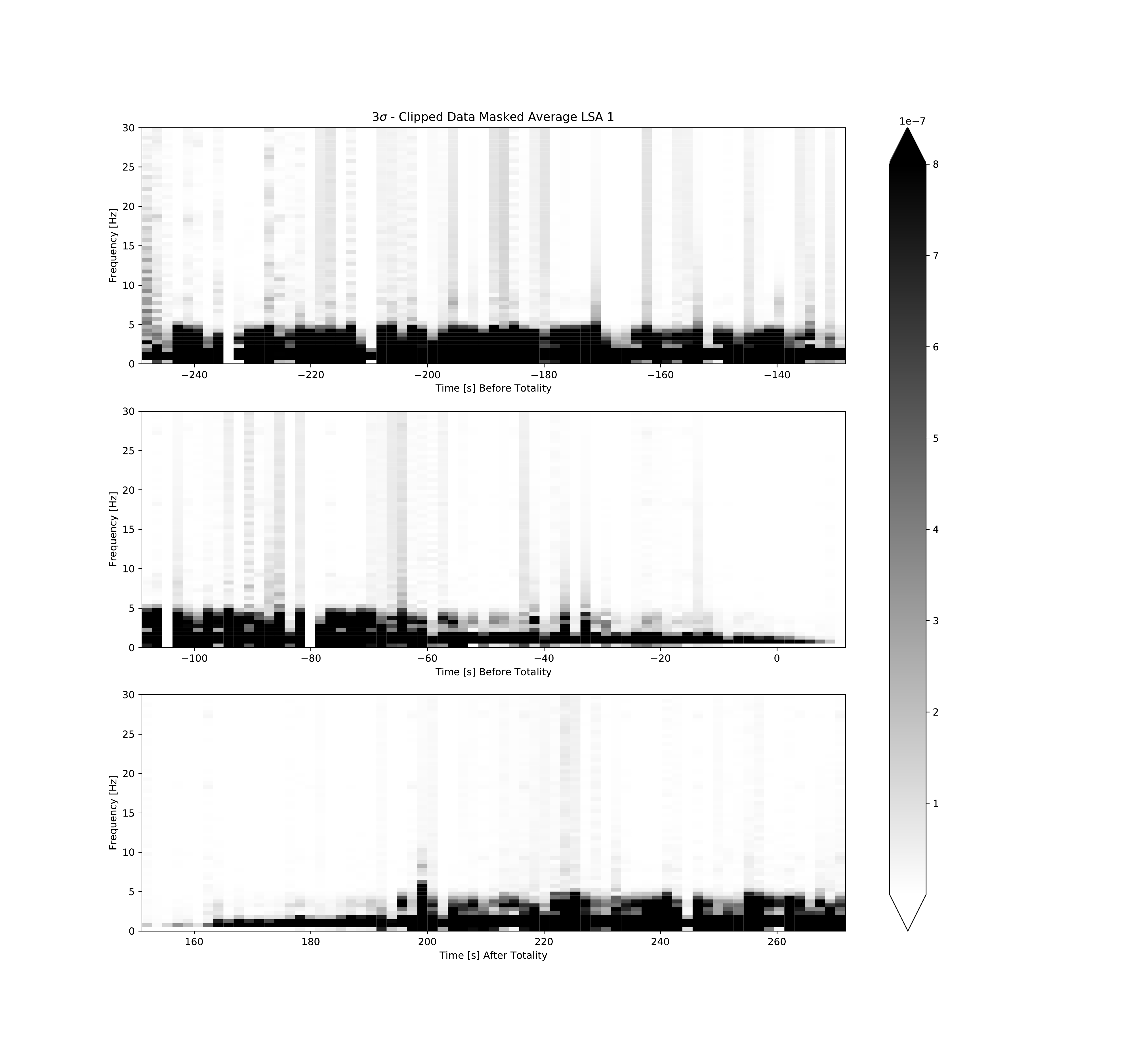}
    \caption{LSA 1 balloon average spectrogram formed from the combined light curves of photodiodes 1, 2, 3, and 4. The spectrograms are shown in three time sections: about 4 to 2 minutes before totality, about 2 before totality, and about 2 minutes after totality. The sustained signal at $\sim 4.5$ Hz is apparent during all these times but fades away during totality.}
    \label{fig:fig8}
\end{figure*}

On the ground we made a video of the shadow bands on a white screen labeled with a scale. We judged that the bands were roughly parallel to the orientation of the bright thin crescent Sun before and after totality, which would be in a direction that is tangent to the Moon's shadow on Earth. If this $\nu \approx 4.5$ Hz frequency signal is identified with the observed peak-to-peak shadow band wavelength of $\lambda \approx 13$ cm (fluctuating between 10 to 15 cm) measured in the video we obtained, it can be inferred that the propagation velocity of the bands ($v = \lambda\nu$) is about $v \approx 59$ cm s$^{-1}$ ($\approx 2.1$ km hr$^{-1}$). 

Finally, some of our ground arrays utilized red and blue filters (see Figure \ref{fig:fig7}), but neither the sustained $\sim 4.5$ Hz signal nor the chaotic higher frequency signals showed a strong dependence on filter color.

\subsubsection{Results from Balloon LSA 1}

Figure \ref{fig:fig8} shows the balloon LSA 1 average spectrogram, which can be directly compared to its ground control counter part LSA 2 in Figures \ref{fig:fig5} and \ref{fig:fig6}. The most obvious result one can glean from Figure \ref{fig:fig8} is that the chaotic signals at frequencies above 5 Hz are absent, making it unlike all ground arrays, but the frequency at $\sim 4.5$ Hz is present, with the same character as seen in all ground arrays. For example, the $\sim 4.5$ Hz signal fades during totality but it reappears after totality. Figure \ref{fig:fig8} only shows signals below 30 Hz, but it should be noted that there is not a 60 Hz signal in LSA 1 since the circuit was powered by battery. 

This is consistent with the idea that the chaotic signals seen on the ground above frequencies of $5$ Hz are due to atmospheric scintillation, but the signal at $\sim 4.5$ Hz is due to a shadow band signal originating outside of the atmosphere.

\section{Discussion and Conclusion} \label{summary}

During the 21 August 2017 total solar eclipse we undertook a study of shadow bands both on the ground and from a high altitude balloon. The signals we detected were very repeatable and therefore robust. Importantly, the spectrogram analysis of the data collected by ground LSA 2 (e.g., Figure \ref{fig:fig6}) used the same model photodiodes as those flown in balloon LSA 1 (Figure \ref{fig:fig8}). LSA 2 reveals that as the light decreases, the intensity patterns of the shadow bands appear. Their appearance occurs approximately four minutes before totality, they are seen to fade out as the Sun is fully eclipsed and they reappear immediately after totality.  

Between $\sim 5$ Hz to $\sim 30$ Hz, the frequency of the shadow band signals are chaotic, short-lived and most prominent, but they likely extend to higher frequencies (e.g., Figures \ref{fig:fig4} and \ref{fig:fig5}). However, these chaotic signals are not present in the balloon LSA 1 spectrogram. These chaotic signals were present in all 16 of our ground photodiode light curves (4 photodiodes in 4 different LSAs). Some of the photodiodes were equipped with color filters, but we found no characteristics that were strongly color dependent. The chaotic signals on the ground were ubiquitous and mostly incoherent over a length scale of 10 cm.     

At frequencies $<5$ Hz there is a persistent frequency at $\sim 4.5$ Hz seen in all spectrograms. 
We suspect that visual observers in our group who were viewing shadow bands on a white screen on the ground primarily saw the $\sim 4.5$ Hz signal, because the higher frequency shadow band signals were very transient and chaotic. As explained in \S5.2.2, we deduced that the bands were oriented in a direction that is tangent to the Moon's shadow on Earth. We also inferred the shadow band wavelength to be $\sim 13$ cm with a propagation velocity $\sim 59$ cm s$^{-1}$ ($\sim 2.1$ km hr$^{-1}$). 

We have some recommendations for those interested in future shadow band research. The photodiodes for all our LSAs had a square spacing pattern of 10 cm. The signal we saw at $\sim 4.5$ Hz was clearly coherent over this spacing, but the ubiquitous signals $>5$ Hz were not well correlated over this spacing. Future researchers who want to study shadow bands are encouraged to use a larger number of photodiodes with a range of separations to explore their coherence.

Another improvement future investigators should implement involves using batteries to power ground LSAs. Not doing this gave rise to a 60 Hz signal (plus harmonics) in our data. It would have been better to use batteries, which were used to power LSA 1 on the balloon. Fortunately, our signals were easy to evaluate using spectrograms. Lastly, with regard to the design of our balloon photodiode array, LSA 1, it is somewhat unfortunate that the ropes connecting the sub-payloads caused the individual photodiode signals to be periodically occulted due to payload spinning. In the end these rope occultations were very low frequency, so they had little effect on our data analysis. However, if future researchers attempt observations from a high altitude balloon, they should  use a design which avoids this.
 
It should be emphasized that the primary goal of our program was not to deploy experimental setups to test the significant predictions reached in the theoretical study of eclipse shadow bands presented by Codona (1986) \citep{1986A&A...164..415C}. Indeed, this would have required more sophisticated ground-based setups than we deployed, as well as setups to measure the properties of the atmosphere during the eclipse. Instead our goal was to attempt to detect shadow bands above the Earth's atmosphere. Surprisingly, we did detect a $\sim 4.5$ Hz optical signal above the atmosphere and on the ground, whereas we only detected chaotic higher frequency signals on the ground. This suggests that the shadow band signal has two components, one originating above the atmosphere and one originating due to atmospheric scintillation. 

We should also emphasize that the Codona (1986) \citep{1986A&A...164..415C} atmospheric scintillation theory (\S2) does make detailed predictions which would be testable with the proper experimental setups. Researchers interested in testing his theoretical predictions are referred to his paper. However, determining the unknown turbulent properties and wind speed components in the atmosphere near the time of totality would be a significant challenge, and this would be a requirement to properly test his theory.

With regard to a shadow band component that originates above Earth's atmosphere, more theoretical work is clearly needed (\S3). 

Most importantly, our results should be confirmed. Future researchers should first decide if they want to study shadow bands due to the turbulent atmosphere, for which all signals were very short lived, versus study any steady signal outside the atmosphere (e.g., like the one we saw at $\sim 4.5$ Hz and which persisted for several minutes before and after totality). Clearly the detection of a steady signal above the atmosphere, as well as on the ground, was our new and unexpected result. In order to study such a signal, future researchers should adopt our approach and deploy experimental setups both above the atmosphere and on the ground. 

Deploying a balloon-based setup more complex than the one we used, while easy to conceive of, is problematic, since current rules dictate that the payload should be under 6 pounds. Future researchers should consider designs which are light-weight and linear, extending perpendicularly in two directions in the x-y plane if possible, with sensors separated by multiples of 10 cm. For example, relative to a central sensor at x = y = 0, additional sensors could extend in both the x and y directions with increasing separations relative to the adjacent sensor of 10 cm, 20 cm, 40 cm, 80 cm, etc., until weight limitations made the design unfeasible. If the last separation was 80 cm, this configuration would be 3 meters long end to end and consist of 17 sensors in the x-y plane. Even if the weight limitations could be managed, flying such a configuration would have to be flight-tested for stability. We believe the goal should be to see if our results are reproducible and undertake measurements over the largest scale possible to study the coherence length. Deploying a ground array of sensors in a similar manner should be powered by battery like the balloon-based sensors; it could conceivably be much larger. Therefore, we encourage additional searches for eclipse shadow bands from high altitude balloon flights and on the ground over a range of separations during future total solar eclipses.

\section*{Acknowledgments}

We acknowledge and thank Dr. Jeffery Vipperman from the Swanson School of Engineering at the University of Pittsburgh for allowing the use of his lab to develop and test some of our equipment. We also acknowledge Aimee Everett, who assisted us with other aspects of the project during the eclipse. We are also grateful to the Clark family (unrelated to co-author R. Clark) from Springfield, Tennessee, who permitted us to use their lawn to launch our high altitude balloon. The eclipse-ballooning work was supported by funding from a NASA Pennsylvania Space Grant Consortium (PSGC) supplemental grant and by cost-share funding from the Dietrich School of Arts and Sciences at the University of Pittsburgh. The opportunity to do this work arose from a project organized by a high altitude balloon team at Montana State University (see \href{http://eclipse.montana.edu}{eclipse.montana.edu}). The project permitted participants to develop their own payloads to fly during the 21 August 2017 total solar eclipse. This work would not have been possible without NASA and Dietrich School funding and the high altitude ballooning expertise provided to us by the faculty, staff, and students at Montana State University.

\bibliographystyle{cas-model2-names}

\bibliography{shadowbands}

\end{document}